\documentstyle[psfig,12pt]{article}
\def\reference{\parskip 0pt\par\noindent\hangindent 0.5 truecm}

\begin{document}

\def\hubh{$H_0=100 h^{-1}$ km s$^{-1}$ Mpc$^{-1}$, $q_0 = 0.1$, $1 \leq h \leq 2$}
\def\gsim{\stackrel{>}{_\sim}}
\def\lsim{\stackrel{<}{_\sim}}

%
% Title
% Capitalise the title normally - do not use ALL CAPS.
%
%\title{Intraday Variability from the Northern Hemisphere}
\title{Intraday Variability in Northern Hemisphere Radio Sources}
%

% Authors
% Here comes the author(s) of the paper. Please add the appropriate author
% names for your paper and indicate within the $^...$ the number(s)
% which corresponds to the institute(s) of each author. In this example
% the second author has two institutional affiliations.
% Add or remove authors as required.
% **** IMPORTANT: Leave the closing curly bracket line as is. ******

\author{T.P. Krichbaum $^{1}$, 
 A. Kraus $^{1}$,
 L. Fuhrmann $^{1}$,
 G. Cim\`{o} $^{1}$,
 A. Witzel $^{1}$ 
} % IMPORTANT: leave this curly bracket as the first character of this line.

% Date - leave this blank.
\date{}
\maketitle

% Institutions
% Here fill in your institute name(s) and address(es)
% The number in $^...$ indicates the author number.  For example
{\center
$^1$ Max-Planck-Institut f\"ur Radioastronomie, Auf dem H\"ugel 69, 53121 Bonn, Germany \\ 
e-mail: tkrichbaum@mpifr-bonn.mpg.de\\[3mm]
}

% Abstract
% Simply place your abstract between the \begin{abstract} and
% \end{abstract} commands.
%
\begin{abstract}
% Place the abstract here.
{\small
\noindent
We summarize results from flux density monitoring campaigns performed
with the 100\,m radio-telescope at Effelsberg and the VLA during the past
15\,yrs. We briefly discuss some of the statistical properties from now
more than 40 high declination sources ($\delta \geq 30^\circ$), which show
Intraday Variability (IDV). In general, IDV is more pronounced for sources with flat
radio spectra and compact VLBI structures. For 0917+62, we present new VLBI images, which suggest
that the variability pattern is modified by the occurrence of new jet components.
For 0716+71, we show the first detection of IDV at millimeter wavelengths (32\,GHz).
For the physical interpretation of the IDV phenomenon, a complex 
source and frequency dependent superposition of interstellar scintillation and source intrinsic 
variability should be considered.
}
\end{abstract}

\noindent
{\bf Keywords: 
quasars: general, quasars: individual (0716+71, 0917+62, \\
0954+65), radio continuum: ISM, scattering}

% Place keywords here. Please write all keywords in lower case. PASA uses the
 %standard list of subject 
% headings adopted by The Astrophysical Journal and available from URL:
%   http://www.journals.uchicago.edu/ApJ/keywords_text.html

% A formatting command to add space between the author list and the body
% of the paper when printed. This spacing may be changed as desired.
\bigskip

%
% Body of paper
%

\section{Statistical Properties of IDV}
\begin{figure}[t]
\makebox[15cm][c]{
\psfig{file=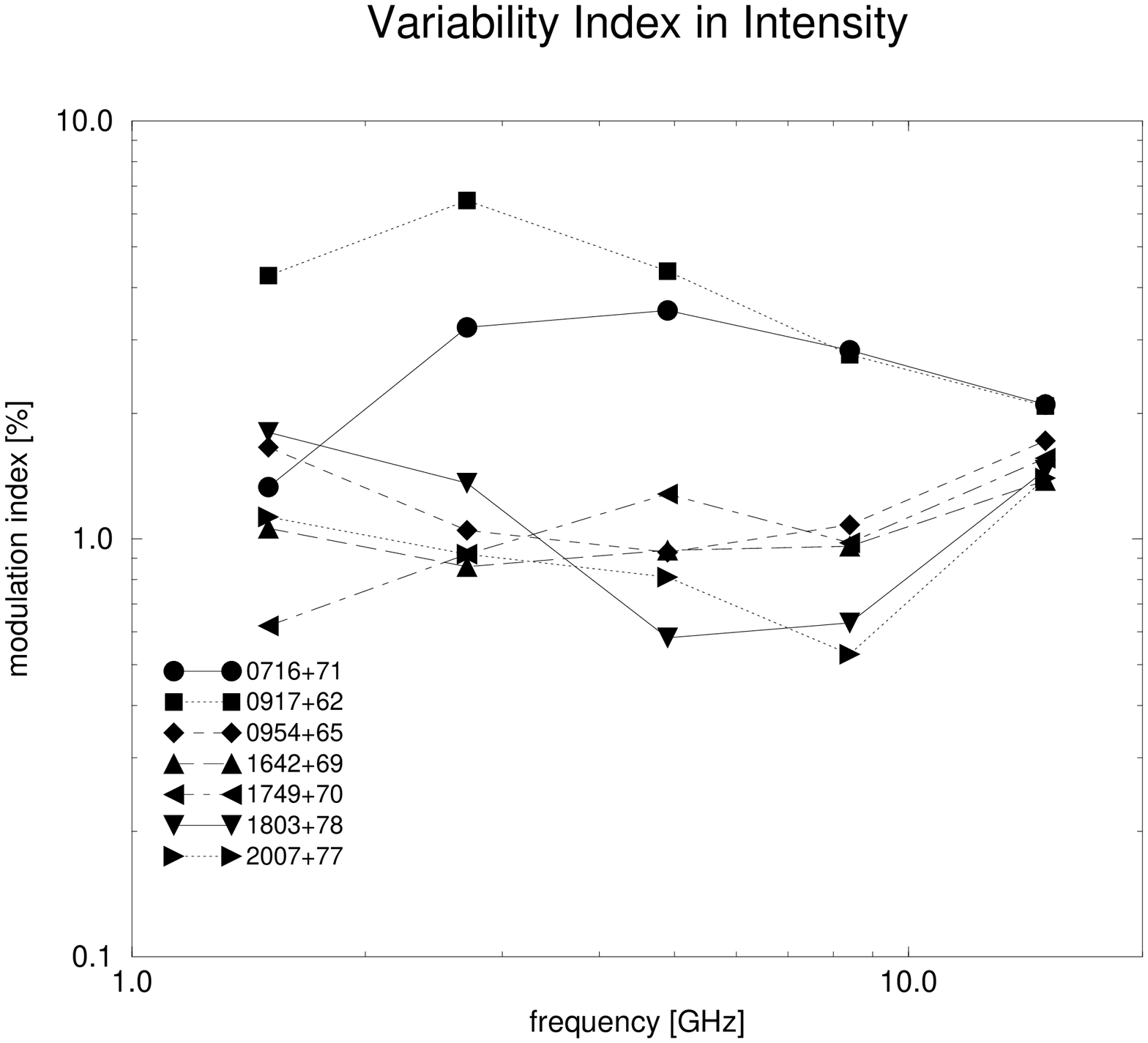,width=7cm,angle=-90}
\psfig{file=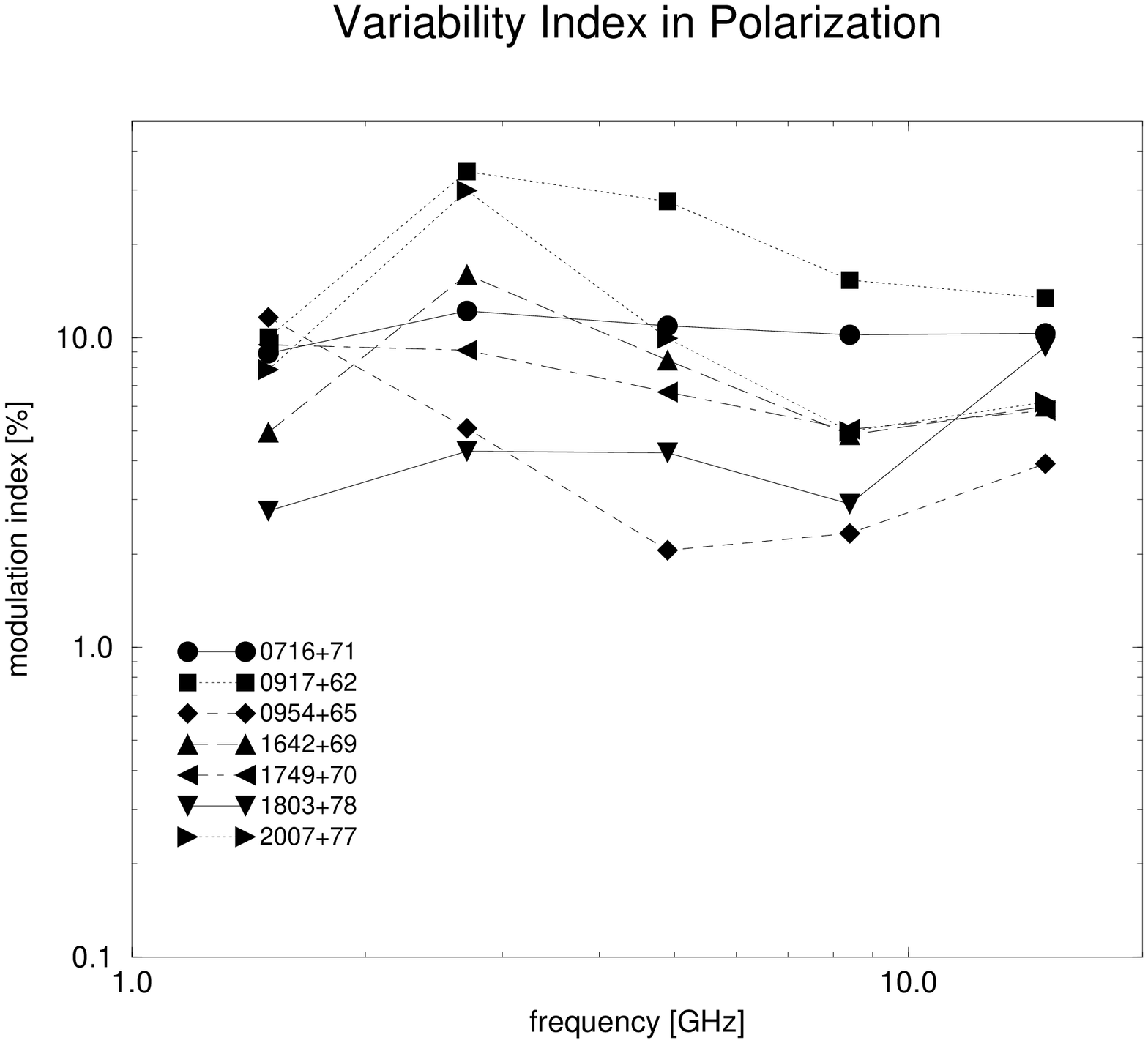,width=7cm,angle=-90}
}
\makebox[15cm][c]{
\psfig{file=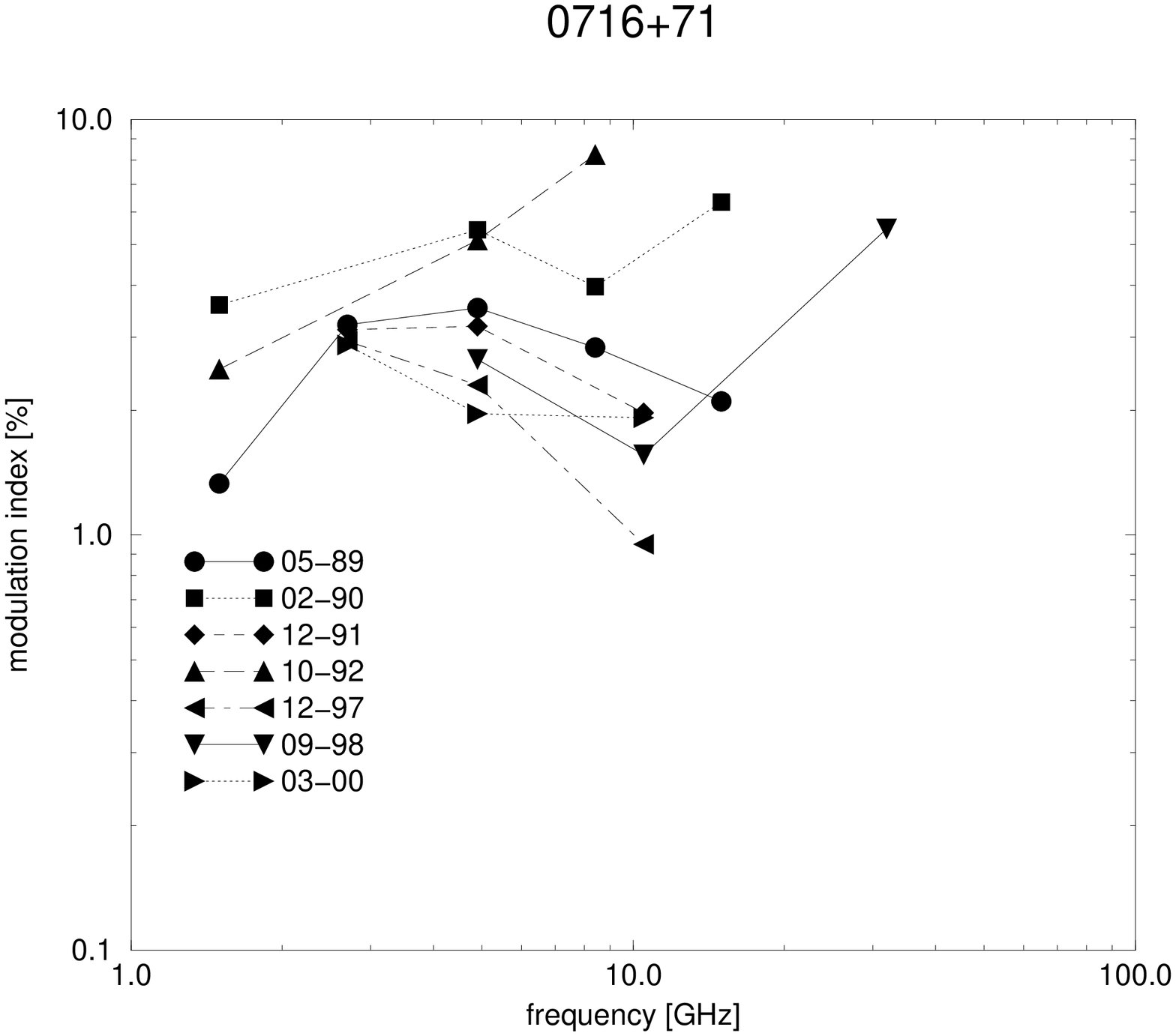,width=7cm,angle=-90}
\psfig{file=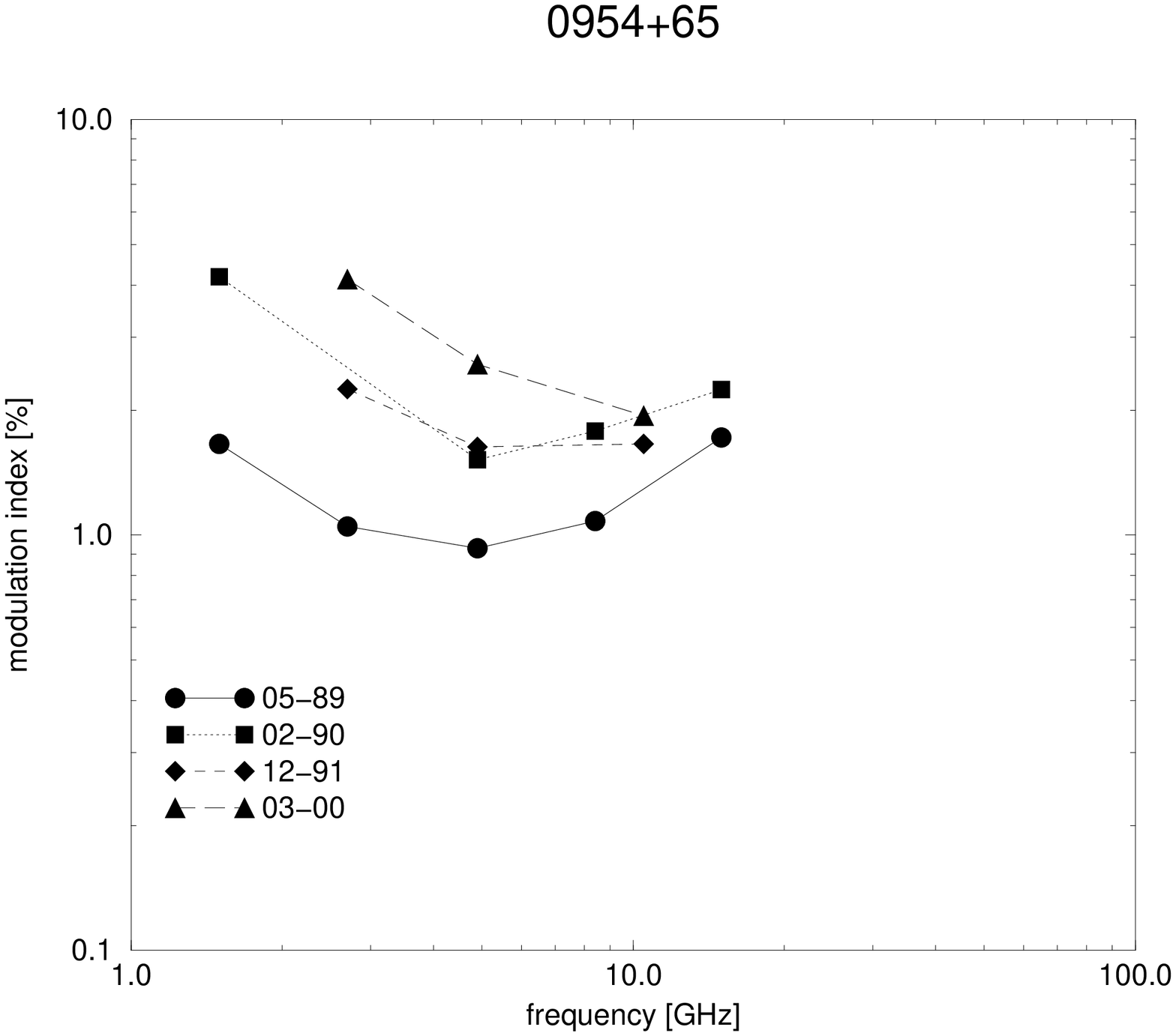,width=7cm,angle=-90}
}
\caption{{\it Top}: Frequency dependence of the variability index $m$ (left: total intensity;
right: polarized intensity) for several IDV sources observed simultaneously with the
VLA and Effelsberg in May 1989.
{\it Below}: The temporal change of the variability index $m$ plotted versus frequency (0716+714 - left,
0954+658 - right) during the last 12\,yrs.
Uncertainties on the individual variability indices $m$ are typically $\Delta m \leq 0.3-0.5$\,\%.
}
\label{mfreq1}
\end{figure}
A large fraction (about 20-30\,\%) of compact flat spectrum radio sources, mostly
identified as Blazars (Quasars, BL\,Lac objects, etc.), show flux density and polarization
variations on timescales of a few hours to a few days. These so called
Intraday Variable (IDV) radio sources (see review of Wagner \& Witzel, 1995 (WW95)) are defined via 
their structure function (SF), exhibiting pronounced maxima within time lags of 
$\sim 0.5-2$\,days (`type II' sources; cf. Heeschen et al. 1987). The Fourier analysis
of the IDV timescales shows in nearly all type II sources a quasi-periodic variability pattern, i.e.
a concentration of the power (in a cleaned power spectrum) to a few ($\lsim 3-5$) discrete
timescales. In contrast, the so called `type I' IDV sources vary slower, showing a quite 
monotonically increasing SF within the time of the observations, suggesting variability timescales
of $\geq 2$ days. Although a $\sim 1$\,day timescale
seems to be common for the classical `type II' IDV sources, `sudden' changes of the variability
timescales are observed eg. in 0716+71, which changed from quasi-periodic daily to less
periodic weekly oscillations (cf. Qian 1995, WW95, Kraus et al. 2001). 
Extremely rapid ($\lsim 0.1$\,days) and pronounced variations 
like the ones recently observed in PKS\,0405-38 (Kedziora-Chudczer et al. 1997) or
J1819+38 (Dennett-Thorpe \& de Bruyn 2000) are not observed in our high declination sample.
The fastest variation observed so far by us, also show variations on a 1\,hr timescale,
however with an amplitude of only 2\,\% (in 0716+71 at 5 GHz in April 1993). At the moment
it is not clear, if J1819+38 or PKS\,0405-38 should be classified as `type II' IDV's, or
if it would be better to define another class of more extreme IDV sources, which would imply that
it is at present unclear whether the physical process causing such extreme variability is
the same as for the `classical' IDV sources.
\begin{figure}[t]
\makebox[15cm][c]{
\psfig{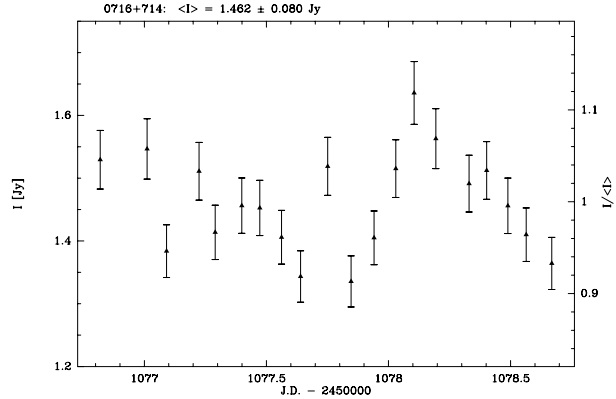}
}
\caption{IDV of 0716+714 at 32\,GHz in September 1998 observed with the
100\,m telescope (Kraus et al. 2001). The peak-to-peak amplitude
is 20\,\%, the typical measurement error is 6\,\%.}
\label{9mm}
\end{figure}

In the radio bands and for classical IDV's, the peak-to-peak variations of
the flux densities can reach up to $20-30$\,\% (eg. for 0804+49, 0917+62), more typically, 
however, are a few percent.
In polarization, the variations are more pronounced ($\sim 20-100$\,\%) and are
about a factor of two faster than in total intensity. The variations in total (I) 
and polarized (P) flux density appear often either correlated (eg. in 0716+71) or anti-correlated
(eg. in 0917+62). However transitions from correlated to anti-correlated variability 
are observed occasionally. Correlated with the variations in P, the polarization angle (PA)
varies on the same timescale by typically a few to a few ten degrees (Kraus et al. 2001). Sometimes, 
polarization angle swings by $180^\circ$ are observed within hours (0917+62: Quirrenbach et al. 1989; 
1150+81: Kochenov \& Gabuzda, 1999). A direct conclusion drawn from the polarization IDV is a multiple
component structure of the radio sources, with their sub-components being of different compactness and
polarization. Although this is in good agreement with the observational
findings from polarization VLBI (cf. Gabuzda et al., this conference), the short 
variability timescales indicate component sizes much smaller than the angular resolution
of present day ground and space-based VLBI ($< 0.2$\,mas). 

\section{Intrinsic or Extrinsic ?}
In Figure 1 (top), we plot the frequency dependence of the variability index $m$ (defined by $m_I=\Delta I/I$
(left) and $m_P=\Delta P/P$ (right), $\Delta I$ and $\Delta P$ are the rms amplitudes) 
for a small sample of sources observed with the VLA and the 100\,m telescope at 5 frequencies 
(1.5 -- 15\,GHz). The observing interval and data sampling restrict
the detection of characteristic variability timescales to $\sim 0.2 \leq t_{\rm var} \leq \sim 2.5$ days
(5 days of data). Still, we use the rms-fluctuation index $m$ to illustrate the strength of variability at
each frequency for {\it these} timescales. We note that for the given (limited) observing interval, $m$ might be
only a coarse estimate of the actual variability pattern, particularly in view of the likely mixture of 
stochastical variations and variations on discrete times scales.
A more detailed structure function analysis is beyond the scope of this paper and 
should be done in the future. Nevertheless, it is obvious that in polarization,
the variability index $m_P$ for most sources peaks somewhere between 1.5 and 5 GHz, in contrast to
the variations in total intensity, where the situation is more complicated. When looking at the
change of the frequency dependence of the variability index with time in Figure 1 (bottom), it is
seen that at least two types of variability exist: in 0716+71 the variability index peaks
for most observing dates between 3-5\,GHz, whereas in 0954+65 a clear minimum of $m$ appears in this
frequency range. Above 8\,GHz and in both sources the variability index increases with frequency!

The theory of interstellar scintillation (ISS) predicts most pronounced variations
close to a critical frequency $\nu_{\rm crit}$, where strong scattering changes into
weak scattering. From Figure 1 it appears, as if this transition frequency is different for
each source and also changes for a given source with time (since the data base is still
small, this effect should be investigated in more detail). Also the strength of variability
at $\nu_{\rm crit}$ seems to be time dependent. An increase of $m$ towards
higher frequencies or a `double peaked' appearance of $m(\nu)$ is inconsistent 
with simple models of ISS, unless a source intrinsic contribution, with increasing
dominance towards higher frequency, is assumed. 

A stratified (multiple layer) or otherwise more 
inhomogeneous interstellar medium may also help to explain the complexity seen in Figure 1. 
As we will show in a moment, this has the potential to explain changes of
$m$ with viewing direction (through our galaxy) and the occurrence of more than one characteristic
variability timescale observed in a few sources (eg. 0716+71, 0917+62), but not the 
observed `sudden' transitions between them\footnote{This should be not confused with the 
annual modulation of $m$ due to the Earth's orbital motion (see Fuhrmann et al., 
this conference; and Rickett et al. 2001).}.  
For refractive ISS (RISS), the variability timescale is $t \propto \theta_{\rm scat} \cdot D / v$, where
$\theta_{\rm scat}=$ scattering size, $D=$ distance to scatterer, $v=$ relative velocity of scatterer.
Rapid changes of $t$ therefore imply similar rapid changes of at least one of the three parameters
in this equation. This could perhaps indicate a very clumpy ISM, eg. with sharp edges of quite fast
moving clouds at different distances. We however think, that this is not very likely. 
If on the other hand, the source size $\theta_{\rm src}$ is of order of $\theta_{\rm scat}$
(`quenched' scattering), a `sudden' ($\leq$ 1 day) change of the characteristic variability 
timescale $t$ implies a similar `sudden' change of the source size 
($t \propto t_{\rm riss} \theta_{\rm src}/ \theta_{\rm scat}$). This, however, is nothing else 
but rapid intrinsic variability. 

Thus, variability amplitudes increasing monotonically with frequency
(from radio, infrared to optical (cf. 2007+77, Peng et al. 2000) and even to X- and Gamma-rays 
(0716+71, Wagner et al. 1996)),
and the sudden transitions of IDV timescales, could also be regarded
as sign for source intrinsic variations, which at least in the case of 0716+71 and 0954+65
appear not unlikely in view of the observed broad-band radio-optical correlations 
(Quirrenbach et al., 1991, Wagner et al. 1993, WW95). The recent finding of IDV in 0716+71 at 9\,mm 
wavelengths (Fig. 2) strongly supports this view. At such short wavelength and high galactic latitude 
($b_{\rm II}=28^\circ$), the 
interpretation of IDV as due to scintillation would appear more questionable. 
More evidence for the idea that IDV at cm-wavelengths is most likely a mixture 
of propagation and source intrinsic effects, also comes from a multi-frequency analysis
of correlated I \& P variations in 0917+62, in which correlated and anti-correlated
I \& P variability peaks were found at 20 and 6\,cm (Qian et al. 2001). Owing to the 
frequency dependence of scattering and 
the fact that for a point source I \& P should vary simultaneously, the correlated peaks were 
interpreted by ISS, resulting in reasonable parameters for the ISM and the source. However, also an
anti-correlated I \& P peak was seen in the same light curve, which -- despite a multiple polarized
component structure assumed for the source -- did not at all fit into the standard scattering model
used by Qian et al.. Since also the polarization angle variations could not be fitted within
this scattering model (adding even more structure components or changing the ISM doesn't help), 
the conclusion must be that (i) either the present ISS model isn't yet developed enough to explain
the polarization angle variations, or (ii) the IDV variations are a mixture of extrinsic and intrinsic 
effects, which only mutually can explain the observed radio light-curves.

\section{VLBI observations of IDV's}
\begin{figure}[t]
%\psfig{file=6cm.ps,width=15cm,angle=-90}
%\vspace{1cm}
\psfig{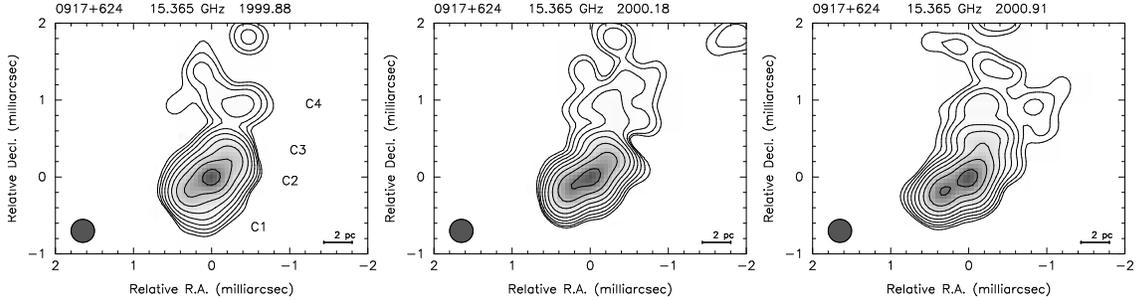}
\caption{The innermost jet region of 0917+62 observed with VLBI at 15 GHz maps during 1999.9--2000.9.
Labels denote identified VLBI components.}
\label{0917maps}
\end{figure}
The extreme apparent brightness temperatures  ($T_B \geq 10^{16}$\,K) deduced from the light travel time
argument require, in this source-intrinsic interpretation of IDV, relativistic 
Doppler-boosting factors $\delta \geq 20$. Recently,
several authors claimed to have seen faster than previously known superluminal motion with speeds
of up to $30 h$\,c (Jorstad et al. 2001, adopting \hubh). 
In 0235+164, a Lorentz-factor
of $\gamma= (30-50) h$ is likely (Qian et al. 2000, Romero et al. 2000). Therefore apparent
brightness temperatures of up to a few times $10^{16} h^3$\,K could in principle be reached with
source intrinsic relativistic jet physics. Much higher brightness temperatures probably indicate 
the presence of scattering or, if intrinsic
mechanisms are at work, an exponent of the Lorentz-transformation for the intrinsic
brightness temperature $T_{B,i}$ larger than 3 ($T_{B,app} \propto \delta^x T_{B,i}$ with $x \geq 3$),
which could appear in non-spherical geometries (Qian et al. 1991 \& 1996, Spada et al. 1999). 
Also other (coherent) radiation mechanisms are not yet ruled out (cf. Benford \& Tzach, 2000,
but see also D. Melrose, this proceeding).

Even if the basic process causing IDV is extrinsic, the  typical properties 
of IDV are affected by source intrinsic variations. A dramatic change in the variability
behaviour of 0917+62, which 
slowed down and nearly stopped to vary in September 1998, but resumed to be rapidly variable 
in February 1999 (Kraus et al. 1999), was recently interpreted as due to orbital motion of the Earth 
relative to the ISM (Rickett et al. 2001, Jauncey \& Macquart, 2001). 
In order to test the suggested annual modulation 
of the IDV pattern, we continued to monitor the source with the 100\,m RT at 
Effelsberg. The results are shown in L. Fuhrmann's paper (this conference), who could not
confirm the annual modulation scenario. 
\begin{figure}[t]
\makebox[15cm][c]{
\psfig{file=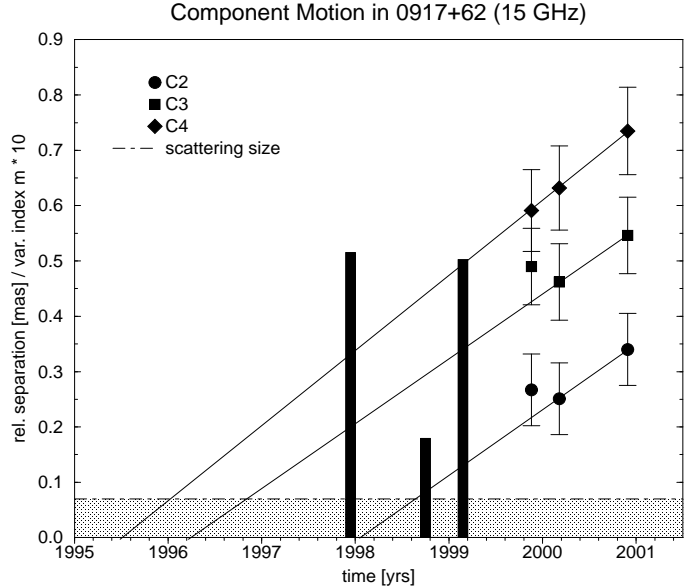,width=9cm,angle=-90}
}
\caption{Relative separation from the VLBI core for the 
components C2, C3 and C4 plotted versus time. Straight lines back-extrapolate the
motion to the time of zero separation. The shaded area indicates a size
smaller than the typical scattering size (of 70\,$\mu$as for 0917+62).
Thick bars show the variability index $m$ (multiplied by a factor of 10)
at the 3 epochs published by Kraus et al. 1999.}
\label{rt}
\end{figure}

In Figure 3, we show some new VLBI maps of the bent milli-arcsecond jet of 0917+62 obtained at
half year time sampling in 1999--2000.
The flux density of the most compact VLBI component C1, 
the `VLBI core' \footnote{`VLBI core' is used in this context synonymous for the transition region at the jet base
from optically thick to thin emission ($\tau=1$--surface), or for the first visible shock near this base.},
is $\sim 200-300$\,mJy (total flux: $\sim 1.5$\,Jy). 
In order to explain the observed variability index of $m \simeq 5$\,\%, 
the core or its associated shock component has to vary with $m= 25 - 40$\,\%. Such large amplitudes are only possible,
if strong rather than weak scattering occurs.

The 2\,cm maps show component motion with
$\beta_{app} = (5-6) h$. Back-extrapolation of the motion of the 3 inner jet components
(C2, C3, C4, Fig. 3) to zero-separation from the stationary assumed VLBI core gives their ejection times
(see Fig. 4). Component C2 was ejected in 1998.1, at a time when 0917+62
showed pronounced IDV. The variation of $m$ with time after ejection of C2 is
indicated by thick bars in Figure 4. The low value of $m$ in September 1998 can 
be interpreted as due to `quenched' scattering and source expansion:
as long as the size of the core region (a blend of core C1 and
moving secondary component C2) is larger than the scattering size,
no IDV is observed. After a while, the secondary component (C2) moves further 
down the jet, expands and fades. The compact VLBI core becomes dominant again and 
the scintillation resumes. 

In order to further test this idea, we searched for additional changes in the variability index of 
0917+62, using IDV observations since 1988. One other particularly low value
of $m=1.4$\,\% in April 1993 coincides nicely with an enhanced 2.8\,cm flux density ($\sim 0.5$\,Jy
excess) and a slightly increased size of the VLBI core at 1.3\,cm at this time (Standke et al. 1996). 
This suggests that also in early 1993 a new jet component was born. Unfortunately 
we have no other high resolution VLBI data for the period 1993--1995 to test this idea. 

\section{Conclusion}

We conclude that the physical interpretation of IntraDay Variability still poses problems,
both to the source extrinsic (ISM) and source intrinsic (AGN related) scenarios.
Most likely, the IDV observed in the radio regime is a quite complex mixture of both effects. 
The relative contribution of these two effects seem to depend on the properties of the ISM,
the viewing direction, and the source size and internal sub-structure, which itself is frequency and
time dependent. There is, however, hope that with multifrequency
variability studies (in intensity and polarization, cf. Fig. 1) and simultaneous VLBI 
monitoring (cf. Fig. 4), the extrinsic and intrinsic contribution to IDV can be separated.

For 0917+62, we conclude that temporal variations of the source size, which are correlated
with total flux density variations and 
the ejection of new VLBI components, are likely to modify the observed scintillation pattern
of compact radio sources, depending on the ratio of actual source size and
scattering size. 

In many compact radio sources, new jet components
(and flux density outbursts observed at high frequencies) appear quite regularly at
one half to a few years time intervals. It is therefore possible to observe an annual modulation in the 
variability pattern (caused by orbital motion) only in those objects, which are quite inactive
and whose nuclear sizes remain smaller than the scattering size for periods much longer than one year.

\section*{References}
\reference Benford, G., \& Tzach, D., MN, 317, 497.
\reference Dennett-Thorpe, J., \& de Bruyn, A.G., 2000, ApJ, 525, 65.
\reference Kochenov, P.Y., \& Gabuzda, D.C.,  1999, in: {\it `BL\,Lac Phenomenon'}, eds. 
 L.O. Takalo \& A. Sillanpa\"a, ASP Conf. Ser. Vol. 159, p. 460.
\reference Heeschen, D.S., et al. 1987, AJ, 94, 1493.
\reference Jorstad, S.G.,  et al. 2001, ApJS, 134, 181.
\reference Kedziora-Chudczer, L.L.,  et al. 1997, ApJ, 490, L9.
\reference Kraus, A. et al. 1999, AA, 352, 107.
\reference Kraus, A., et al. 2001, AA, submitted.
\reference Jauncey, D.L., Macquart, J.P., 2001, AA, 370, 9.
\reference Peng, B., et al. 2000, AA, 353, 937.
\reference Qian S.J., et al., 1996, in: {\it `Energy Transport in Radio Galaxies and Quasars'},
 eds: P. Hardee et al., ASP Conf. Ser. Vol. 100, p. 55.
\reference Qian, S.J., 1995, Chin. Astron. Astrophys. 19/1, p.~69.
\reference Qian, S.J., et al. 2000, AA, 357, 84.
\reference Qian, S.J., et al. 2001, AA, 367, 770.
\reference Quirrenbach, A., et al. 1989, AA, 226, 1.
\reference Quirrenbach, A., et al. 1991, AA, 372, 71.
\reference Rickett, B.J., et al., 2001, ApJ, 550, 11.
\reference Romero, G.E., Cellone, S.A., \& Combi, J.A., 2000, AA, 360, 47.
\reference Spada, M., Salvati, M., Pacini, F., 1999, ApJ, 511, 126.
\reference Standke, K., et al., 1996, AA, 306, 27.
\reference Wagner, S.J. et al., 1993, AA, 271, 344.
\reference Wagner, S.J. et al., 1996, AJ, 111, 2187.
\reference Wagner, S.J., \& Witzel, A.,  1995, ARA\&A, 33, 163 (WW95).

\end{document}